\documentclass[aps,prl,preprint,showpacs]{revtex4}
\usepackage{amsmath}
\usepackage{amssymb}
\usepackage{graphicx}% Include figure files
\usepackage{subfigure}
\usepackage{dcolumn}% Align table columns on decimal point
\usepackage{bm}% bold math
\begin{document}

\title{Trading leads to scale-free self-organization} 

\author{M. Ebert} \author{W. Paul \email{Wolfgang.Paul@Uni-Mainz.De}}
\affiliation{Department of Physics, Johannes-Gutenberg University,
  55099 Mainz, Germany}

\date{\today}

\begin{abstract}
Financial markets display scale-free behavior in many different
aspects. The power-law behavior of part of the distribution of
individual wealth has been recognized by Pareto as early as the
nineteenth century. Heavy-tailed and scale-free behavior of the
distribution of returns of different financial assets have been
confirmed in a series of works. The existence of a Pareto-like
distribution of the wealth of market participants has been connected
with the scale-free distribution of trading volumes and
price-returns. The origin of the Pareto-like wealth distribution,
however, remained obscure. Here we show that it is the process of
trading itself that under two mild assumptions spontaneously leads to
a self-organization of the market with a Pareto-like wealth
distribution for the market participants and at 
the same time to a scale-free behavior of return fluctuations. These
assumptions are  (i) everybody trades proportional to his current
capacity and (ii) supply and demand determine the relative value of
the goods.  

\end{abstract}
\maketitle

Exchange of goods, i.e. trading, is a process as old as humanity. We
show here that under two rather mild assumptions trading always leads
to a Pareto-type wealth distribution\cite{pareto} and the non-Gaussian price
fluctuations \cite{mandelbrot,ms-nature-95,stanley-new} that have
attracted the interest of statistical 
physicists \cite{stanley,bouchaud}. Price fluctuations of financial assets display
heavy-tailed non-Gaussian behavior over a broad range of time scales
before becoming Gaussian as required by the central limit
theorem. Their distribution has been described by truncated L\'evy
distributions\cite{tr-levy} as well as having a power-law tail with an exponent
around $x=-4$ outside the range of L\'evy-like power law tails \cite{stan-powerlaw}. Both
descriptions, however, contain scale-invariant power-law distributions
of the price fluctuations over a limited range in size and over a
limited time horizon. Another well-known and famous power-law 
occurring in our societies is the Pareto law of wealth
distribution\cite{pareto}. This law has not only been found for the wealth of
individuals \cite{unu} but also for the size of companies \cite{wealth-comp}.  

We will show that the simplest imaginable model of trading, where
trading decisions are made randomly, where always a finite fraction of
available goods is invested and where the imbalance between supply and
demand changes the price of goods, is sufficient to produce a Pareto
type wealth distribution as well as scale-free fluctuations of the
market. In fact, independent of the choice of parameters, our model
market always self-organizes into a stationary, scale-free state. Only
the duration of the transient, and parameters quantifying the
transient and the stationary state are influenced by the choice of the
model parameters. 

Agent based models are commonly used to analyze markets. But often the
complexity of these models obscures the cause of the observed market
behavior. We are therefore looking at a minimal model for agent based
trading. Our model simulates the exchange of two goods. We will call
them stock and money from here on but this denomination is arbitrary
and only serves to facilitate the description of the results. Each
agent at each time step randomly decides whether he wants to buy or
sell stock. The quantity of money spent on buying stock, or stock sold
for money is also determined randomly, varying between nothing and an
investment fraction $c$ of the agents current posession in stock or
money, respectively. 

\begin{figure}[htb]
\vspace*{5mm}
\begin{center}
\includegraphics[width=0.7\columnwidth,angle=-90]{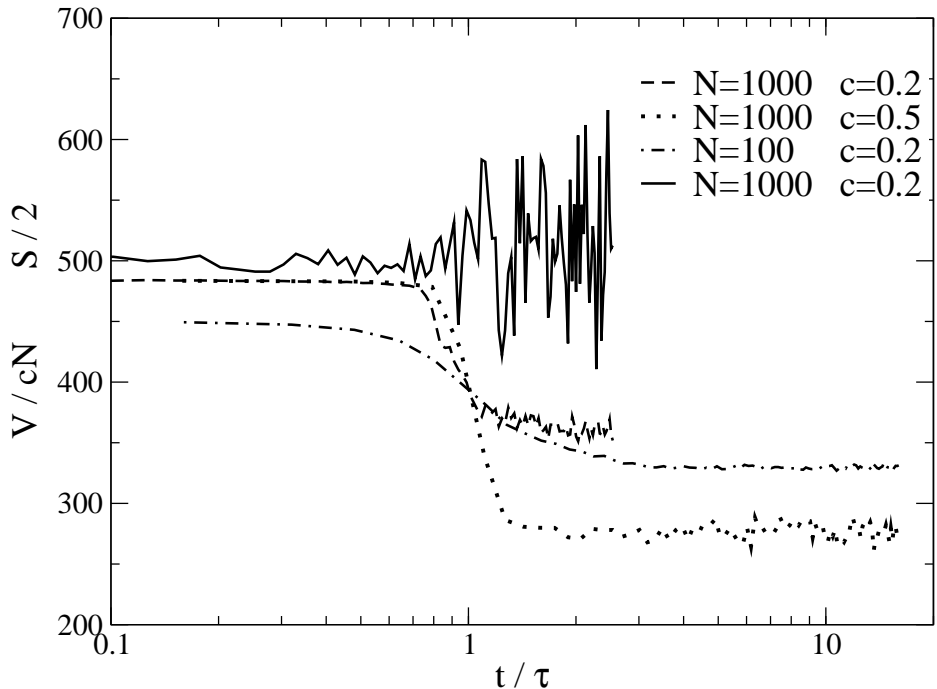}
\caption{Trading volume per time step normalized to its expected average
value $1000 Nc/2$ for several choices of numer of agents $N$ and
investment fraction $c$ (broken curves) and price development for one of
these cases (full curve) as a function of scaled time. The time
scaling $\tau$ is chosen as the center of the crossover from  high trading
volume to low trading volume.} 
\label{fig1}
\end{center}
\end{figure}
Agents are not allowed to leave or enter the market and neither stock
nor money may be created or destroyed. The price is determined by
supply $S$ and demand $D$. To capture the conviction that supply and
demand determine the price one could use a simple ansatz  to adjust
the price $p$ in each time step according to $p(n+1) = p(n) D / S $. To
avoid the singularity for $S=0$ we instead used   
\begin{equation}
p(n+1) = p(n) [1 + (D - S) / (D + 2S)] = p(n) [1 + f(S,D)]
\label{supdem}
\end{equation}
This formula limits price fluctuations in each time step to a factor
of $2$, is differentiable and has no singularities. Several
possibilities for price changes have been analyzed, however, and we
found that our model is not sensitive to the choice of price update
function, as long as it guarantees a price adjustment monotonously
depending on the imbalance between supply and demand.  
We can write down the master equation for the probability $P(A, G, n+1)$
for an agent to possess a number $A$ of stocks and an amount $G$ in money
at time step $n+1$: 
\begin{eqnarray}
&P\left(A, G, n+1 \right) = (1 - q(n)) P\left(A, G, n\right)\\
&+ \frac{q(n)}{2c} \sum_{A' = A - \frac{cG'}{p(n)}}^{A-1}
\frac{P\left( A', G + (A'-A) p(n), n \right)}{G + (A'-A) p(n)} \nonumber\\ 
&+ \frac{q(n)}{2c} \sum_{A' = A+1}^{A / (1-c)} \frac{P\left( A',
G + (A'-A) p(n), n \right)}{A'} \nonumber
\label{eq:masterfinanz}
\end{eqnarray}
with 
\begin{equation}
q(n)
= 1 - \left| \frac{ \sum_i x_i(n) \left( A_i(n) - \frac{G_i(n)}{p(n)}
  \right) }{\sum_j x_j(n) \left( A_j(n) + \frac{G_j(n)}{p(n)} \right) }
\right| 
\label{eq:akzeptanz}
\end{equation}
being the propability of a trade attempt actually happening. $x_i(n)$ is
the fraction of stock or money of agent $i$ traded in the step, randomly
chosen between $0$ and $c$. We begin our simulations in a perfect
socialistic state: all agents possess the same wealth, e.g. $1000$
stocks worth $1000$ units and $1000000$ units of money. 
The time evolution of trading volume and stock price generated by our
market model are shown in Fig.~\ref{fig1}. For several choices of the number of
participating agents $N$ and investment fraction $c$, Fig.~\ref{fig1} shows that
there is a crossover between two dynamic regimes. Between the two
regimes the trading volume decreases and simultaneously the average
price increases and also the price fluctuations increase at a
crossover time $\tau$ that scales as $(N/c)2$ (Fig.~\ref{fig1}
displays this crossover in scaled time).  

\begin{figure}[htb]
\vspace*{5mm}
\begin{center}
\includegraphics[width=0.7\columnwidth]{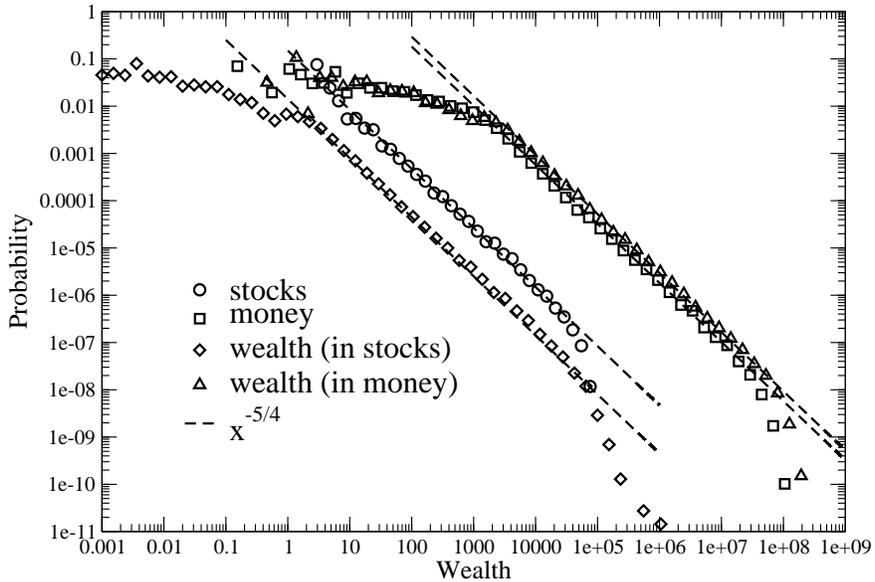}
\caption{Probability density for stocks owned (circles), money owned
(squares), total wealth measured in stocks (diamonds) and total wealth
measured in money (triangles) for $N=200$ agents and investment fraction
$c=0.5$.} 
\label{fig2}
\end{center}
\end{figure}
The initial regime is characterized by Gaussian fluctations. The
initial $\delta$-distribution of wealth spreads into a Gaussian distribution
with increasing width. When the influence of the boundary at zero
wealth starts to be felt, the wealth distribution changes into a
lognormal form with a center which moves towards zero wealth. If
growth of wealth would be strictly proportional to existing wealth \cite{gibrat}
this would be a stationary state of the wealth
distribution. Furthermore, in this early time regime the price
fluctuates around an equilibrium value which is simply given by the
available money per stock. If one does not start the simulation with
this value, the price adjusts to it within a few time steps. Price
fluctuations are Gaussian and the distribution of trading volumes in
one time step is Gaussian as well. 

However, this regime is not dynamically stable. The wealth
distribution develops heavy tails and finally crosses over to its
stationary shape which displays power law behavior. This is shown in
Fig.~\ref{fig2} which shows several measures of the wealth per agent in the
stationary state for a system with $N=200$ agents which invest at most
$50$~\% of their wealth in each step. In the stationary regime, the
distribution of stocks per agent is a perfect power law with an
exponent $x=-1.25$, i.e., the trading process leads to a
self-organization of the market into a scale-free state. Varying
the number of agents between $N=100$ and $N=1000$ and the investment
fraction between $c=0.2$ and $c=0.9$ we always find a power law
behavior with exponent around $x \simeq -1.3$. The distribution of money
owned is almost constant for small amounts of money, then crosses
over to the same power law found for the number of stocks owned
and finally, for large amounts of money, a steeper decay is found
which reflects the finiteness of the money present in the
market. To determine the total wealth of an individual, we can
either express wealth in units of stocks or express it in units of
money. The money value of an amount of $A(n)$ stocks is $G(n) = p(n)
A(n)$, and the stock value of an amount of money $G(n)$ is $A(n) =
G(n) / p(n)$, where $p(n)$ is the current price of stocks on the
market. Thus total wealth always  depends on the price dynamics
and this enters asymmetrically between stocks and money. However,
for the case of $c=0.5$ shown in the Fig.~\ref{fig2} both ways of measuring
wealth show power law regimes over $3-4$ orders of magnitude with
the same epxonent of $x=-1.25$ as the other distributions. The price
fluctations only shift the wealth distribution expressed in money
to slightly larger wealth values. This is changed for large $c$
where both distributions show a differing power law exponent due
to the qualitative change in price fluctuations observed then (see
Fig.~\ref{fig4}). Although the exponent found for the scale invariant part
of the wealth distribution differs from the typical value $x_P \simeq 2.5$
found for example for a recent study on the world distribution of
household wealth \cite{unu}, our very simplified economy is able
to capture the emergence of a Pareto law. It is also interesting to note,
that in the scale-free state, the average stock price is larger
than the average money available per stock.  

\begin{figure}[htb]
\vspace*{15mm}
\begin{center}
\includegraphics[width=0.7\columnwidth]{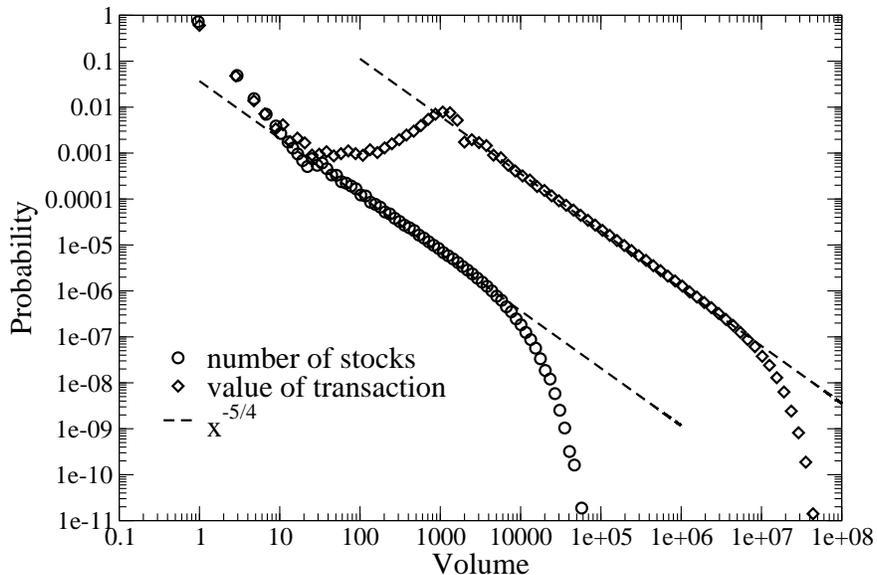}
\caption{Probability density for the trading volume per transaction and
money invested per ransaction for an investment fraction of $c=0.5$.} 
\label{fig3}
\end{center}
\end{figure}
There have been several attempts in recent years to explain the
origins of the Pareto distribution of wealth by assuming interactions
between agents to resemble collisions between particles, as treated
using kinetic theory in statistical physics. Often, the exchange of
wealth in these binary collisions is modelled by a multiplicative
stochastic process \cite{solomon}. If $W_i(n)$ is the wealth of agent
$i$ at time $n$, one writes $W_i(n+1) = W_i(n)(1+\epsilon \xi(n))$
where $\epsilon << 1$ is a small amplitude and 
$\xi(n)$ a random number in $[0,1)$. The stationary distribution of wealth
is then scale-free, generally with an exponent $x=-2$ \cite{chak-2003}, also not
in agreement with the empirical value $x_P$. Introduction of
heterogeneous behavior among the agents, e.g., in the form of a
random saving propensity \cite{chak-2007,germ-2006} can change the observed exponent to
the empirical value. The generic behavior of these kinetic models
for wealth redistribution has also been worked out \cite{duering,lammoglia}.  

In contrast to this approach, in our model wealth changes in a trade
only due to the update of the price of goods. The wealth change of an
agent is asymmetric between buyers and sellers 
\begin{eqnarray*}
W_i(n+1) &=& W_i(n) [1+ f(S,D)(1-c \xi(n))]\\ 
&-& f(S,D)(1- \xi(n))G_i(n)\quad
\mbox{(seller)}\\
W_i(n+1) &=& W_i(n)[1+ f(S,D)c\xi(n)]\\
&-& f(S,D)(1-\xi(n))A_i(n)p(n)\quad      \mbox{(buyer)} 
\end{eqnarray*} 
where $\xi(n)$ is a uniform random number in $[0,1)$. The trading process
itself (i.e., the collision between the particles in the physical
analogy) does not change the wealth of the agents, as they only
exchange stocks for money and vice versa. Only the interaction
between trading process and price update leads to a change of wealth
and the observed self-organization of the wealth distribution into a
Pareto-like state. For this to occur one has to have at least two
goods which are traded, one of which is conveniently identified as
money. This simplest possible economy thus inevitably leads to a
Pareto-like behavior for large wealths. 

%\begin{figure}[htb]
%\vspace*{5mm}
%\begin{center}
%\includegraphics[width=0.7\columnwidth]{fig4.eps}
%\caption{Probability density for the trading volume per Monte Carlo
%step for three choices of investment fraction. Lines are lognormal
%fits to the distributions. } 
%\label{fig4}
%\end{center}
%\end{figure}

We argued that it is the coupling of trading to the price process that
leads to the observed power-law behavior. So let us now look at the
stationary behavior for the price fluctuations. Gabaix et al. \cite{stanley-nature-2003}
presented a model which linked the presence of a Pareto-type
distribution of the wealth of market participants to the occurrence of
power law tails in the distribution of trading volumes and price
fluctuations. The exponents characterizing these power laws $p(|r_t| >
z) \propto z^{-\beta}$ and $p(V_t > u) \propto u^{-\gamma}$ where
$r_t=ln[p(t_0+t) / p(t)]$ is the return 
on a time horizon $t$ and $V_t$ is the volume traded on time horizon $t$, are
related through the price impact function. Several empirical studies
of different markets cite{volume-impact,farmer,zhou} have established that over a reasonable
range in volumes, price impact and volume are related through a power
law dependence $\Delta p \propto V^\alpha$. From this it follows that
$\beta=\alpha\gamma$. The observed 
exponents are $\alpha \simeq 1/2$ \cite{volume-impact,farmer},  $\alpha \simeq 0.2$ for
larger volumes \cite{farmer} and  $\alpha \simeq 0.6 - 0.7$ \cite{zhou}.  

The distribution of volumes for individual trades in our model is
shown in Fig.~\ref{fig3}. Both in terms of number of stocks traded and in terms
of money value of the transaction the volume obeys a power law scaling
with the same exponent as the wealth distribution (i.e. $\alpha =1$) over
three orders of magnitude before the cutoff due to finite total wealth
in the system sets in. For small volumes there seems to exist another
regime which is dominated by the low-end part of the wealth
distribution. Our model therefore conforms to the phenomenological
finding that a power law wealth distribution gives rise to a power law
volume distribution. The price, however, is updated according to the
difference between supply and demand as given by equ.(~\ref{supdem}). The update
function depends on the imbalance between supply and demand, $I=S-D$,
and the cumulative volume $V=min(S,D)$ of all trades in one Monte Carlo
step This distribution does not show a
scale-free regime, displays a maximum at intermediate volumes and can
be phenomenologically described by a lognormal form. The distribution
broadens with increasing investment fraction and becomes highly
assymmetric for large investment fraction. 
For all choices of investment fraction this leads to non-Gaussian
heavy tails in the distribution of price fluctuations as shown in
Fig.~\ref{fig4}. The L\'evy-tails 
found for $c=0.9$ are the origin of the different behavior of the wealth
distribution whether expressed in number of stocks or in money
(discussed for Fig.~\ref{fig2}), since the latter is susceptible to the
L\'evy-type fluctuations in the price. For an intermediate range of
investment fractions, however, the tail of the distribution of price
flucuations is compatible with the behavior phenomenologically found
for real markets.  For $c=0.5$ we even recover the exponent $x=-4$ found
in Ref. \cite{stan-powerlaw}.
  
\begin{figure}[htb]
\vspace*{5mm}
\begin{center}
\includegraphics[width=0.7\columnwidth]{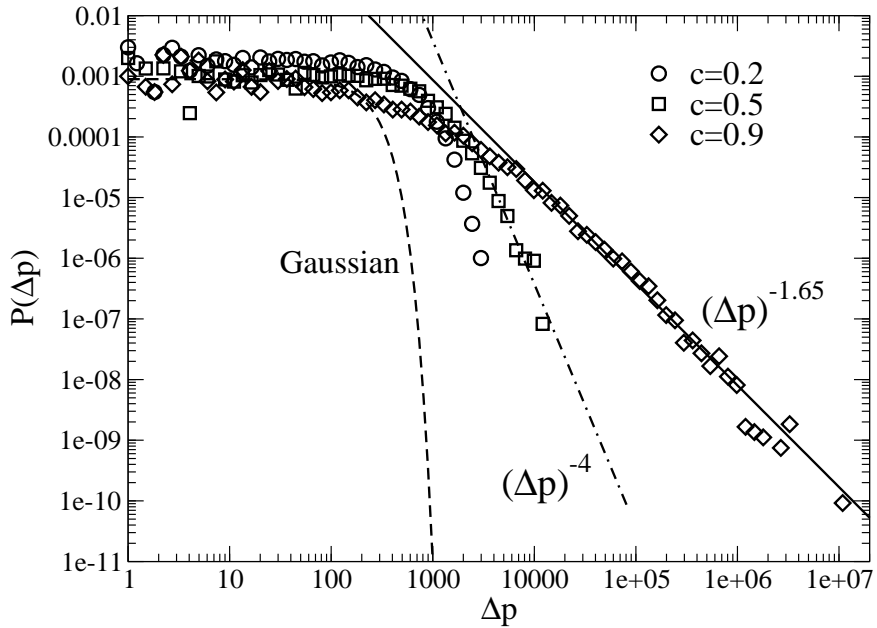}
\caption{Probability density for price fluctions in one Monte Carlo
step. For all choices of investment fraction c we obtain distributions
with a fat tail compared to the Gaussian behavior. For small and
intermediate c the tail behavior is compatible with the market
phenomenology, for $c=0.9$ we obtain a L\'evy-type fat tail.} 
\label{fig4}
\end{center}
\end{figure}
The market behavior discussed above remains qualitatively unchanged
when we include interest in our model (i.e., increase the money
available to the agents by a fixed rate), the price however, now
fluctuates around an increasing average. We would like to note also
that our model remains strictly egalitarian. There is no symmetry
breaking between the different agents and each agent wanders up and
down the wealth curve in the course of the simulation (which is, of
course, not very realistic). This shortcoming and the deviation of the
Pareto exponent from the phenomenologically observed one can be
amendet when one introduces heterogeneous trading strategies among the
agents. One could even solve an inverse problem to determine the
trading strategy leading to the correct exponent \cite{lammoglia}. However, we would
like to emphasize here that already the extremely simplified model we
introduced captures the emergence of the Pareto law and the
qualitative relations between wealth distribution and market
fluctuations. 

The self-organization of the market into a scale-free wealth state and 
the connection between the wealth distribution and the price
fluctuations found at the market lends 
itself to a consideration of the effect that leverage has in such a
situation. The importance of leverage to generate the Pareto-tail of
the wealth distribution has been discussed long ago by Montroll and
Shlesinger \cite{montroll}. Leverage virtually increases the amount of money present
in the system and with that the point to which the Pareto-like wealth
distribution extends before it is cut off. The extend of the Pareto
distribution in turn, determines the scale of the price fluctuations,
Employing leverage thus increases the scale of the fluctuations present in the
market. When too much leverage is employed this creates
downward-fluctuations exceeding the real existing wealth of the
agents, arguably leading to the credit defaults the financial markets
are plagued with at the moment.  
Our simple model also suggests that economic inequality \cite{gibrat} is almost as
old as humankind. When the first trading good was invented that could
be used in the way money is used today, and was not meant to be
directly consumed or used up in other ways in daily life, fluctuations
in the exchange rate of this good with others destabilized the state
of economic equality and led to a Pareto-type wealth distribution.

{\bf Acknowledgement:} The authors thank J. J. Schneider, T. Preis and
J. Zausch for discussions.

%\clearpage

\newpage
\end{document}